\documentclass[prl,epsfig,twocolumn,showpacs,preprintnumbers,amsmath,amssymb]{revtex4}
\usepackage{amsmath}
\usepackage{epsfig}
\usepackage{graphics}
\begin{document}
\title{Frequency and damping of the Scissors Mode of a Fermi gas }

\author{G.\ M.\ Bruun and H.\ Smith}

\affiliation{Niels Bohr Institute, University of Copenhagen,
Universitetsparken 5, DK-2100 Copenhagen \O, Denmark.}

\date{\today{}}

\begin{abstract}
We calculate the frequency and damping of the scissors mode in a
classical gas as a function of temperature and coupling strength.
Our results show good agreement with the main features observed in
recent measurements of the scissors mode in an ultracold gas of
$^6$Li atoms. The comparison between theory and experiment
involves no fitting parameters and thus allows an identification
of non-classical effects at and near the unitarity limit.
\end{abstract}


\maketitle

\section{Introduction}
By tuning the atom--atom interaction in a gas of fermions it is
possible experimentally to investigate the crossover from a
Bose--Einstein condensate (BEC) of molecules to a BCS superfluid. In
the crossover region the system is strongly interacting in the sense
that $k_F|a|\gg1$ (unitarity limit), where $a$ is the $s$-wave
scattering length for the atom--atom interaction and
$k_F^3=3\pi^2n$, with $n$ being the density of the gas. A
theoretical description of this region is challenging, in particular
at non-zero temperature~\cite{Giorgini}.

Collective mode experiments have given a wealth of insight into the
properties of atomic gases, since they often provide more detailed
information than e.g. thermodynamic measurements~\cite{Grimm,Kinast1}.
In a recent experiment~\cite{Wright} the scissors mode excitation
in an elliptical trap was used to characterize the transition
between hydrodynamic and collisionless behavior as a function of
temperature and scattering length. Since both the frequency and
attenuation is affected by the atom--atom interaction, such
measurements can give important information on the properties of a
fermion system at and near the unitarity limit.

In this Brief Report we calculate the frequency and attenuation of
the scissors mode investigated in Ref.~\cite{Wright}, assuming the
temperature to be sufficiently high that the gas can be treated as
being classical. Since the calculation involves no fitting
parameters, it can be used to identify non-classical features of the
observed frequency shift and damping as functions of temperature
and/or interaction strength.

\section{Kinetic theory for a classical gas}
Consider a two-component (for brevity denoted by "spin" with the two
values $\sigma=\uparrow,\downarrow$) Fermi gas of atoms with mass
$m$ in its normal phase trapped in a potential $V({\bf
r})=m(\omega_x^2x^2+\omega_y^2y^2+\omega_z^2z^2)/2$. At high
temperatures, in the classical regime,  the dynamics is described by
a semi-classical distribution function $f({\bf{r}},{\bf{p}},t)$,
which satisfies the Boltzmann equation. For the scissors mode
studied experimentally in \cite{Wright}, the two components of the
gas move together, and we need only consider one distribution
function $f=f_{\uparrow}=f_{\downarrow}$.

To calculate the frequency and damping of the scissors mode, we
linearize the Boltzmann equation in terms of a small deviation
$\delta f=f-f^0$ from the equilibrium distribution
$f^0({\mathbf{r}},{\mathbf{p}})$ by writing $\delta
f({\mathbf{r}},{\mathbf{p}},t)=
f^0({\mathbf{r}},{\mathbf{p}})\Phi({\mathbf{r}},{\mathbf{p}},t)$.
The linearized Boltzmann equation becomes
\begin{equation}
f^0\left(\frac{\partial \Phi}{\partial t} +\dot{\bf
r}\cdot\frac{\partial \Phi}{\partial{\bf r}} +\dot{\bf
p}\cdot\frac{\partial \Phi}{\partial{\bf
p}}\right)=-I[\Phi]\label{Boltzmann},
\end{equation}
where $\dot{\bf r}={\bf v}={\bf p}/m$ and $\dot{\bf p}=-\partial
V/\partial{\bf r}$. The collision integral $I$ is given in Ref.\
\cite{Massignan} with an interaction described by $s$-wave
scattering involving only particles with opposite spin. The cross
section $\sigma$ is
\begin{equation}
\sigma=\frac{4\pi a^2}{1+(p_ra/\hbar)^2}, \label{crosssection}
\end{equation}
where ${\bf p}_r$ is the relative momentum of the scattering
particles. The unitarity limit is defined by
$|a|\rightarrow\infty$.

In the hydrodynamic limit, a scissors mode in the $xy$ plane is
characterized by a velocity field ${\bf v}\propto\nabla (xy)$ or
\begin{equation} {\bf v}=b(y,x,0),\label{velfield}\end{equation} where $b$
is a constant. To describe this mode we choose an ansatz for $\delta
f$ of the form
\begin{equation}
\Phi({\mathbf{r}},{\mathbf{p}},t)=(c_1xy+c_2xp_y+c_3yp_x+c_4p_xp_y)e^{-i\omega
t},
\end{equation}
where the $c_i$ are constants and $\omega$ is the mode frequency.
We insert this ansatz into the linearized Boltzmann equation
(\ref{Boltzmann}) and take moments by multiplying by any of the
terms $xy,xp_y,yp_x$ and $p_xp_y$ appearing in $\Phi$ and
subsequently  integrating over both $\mathbf r$ and $\mathbf p$.
The result is a homogeneous  set of four coupled equations for the
coefficients $c_1,  \ldots , c_4$. The frequencies of the
collective modes are determined by the roots of the corresponding
determinant which yields the equation
 \begin{equation}
 \frac{i\omega}{\tau}(\omega^2-\omega_h^2)+(\omega^2-\omega_{c1}^2)(\omega^2-\omega_{c2}^2)=0.\label{Determinant}
 \end{equation}
Here
\begin{equation}
\omega_h=\sqrt{\omega_x^2+\omega_y^2}
\end{equation}
is the mode frequency in the hydrodynamic limit and
\begin{equation}
\omega_{c1}= \omega_x+\omega_y \;\;{\rm and}\;\; \omega_{c2}=
|\omega_x-\omega_y|
\end{equation}
the mode frequencies in the collisionless limit~\cite{Odelin}. The
viscous relaxation rate in (\ref{Determinant}) is given by
\begin{equation}
\frac{1}{\tau}=\frac{\int d^3{r}d^3{p}p_xp_yI[p_xp_y]}{\int
d^3{r}d^3{p}p_x^2p_y^2f^0}.
\end{equation}
In the classical limit, one obtains~\cite{Bruun72}
\begin{equation}
\frac{1}{\tau}=\frac{N}{5\pi^2}\frac{m\bar{\omega}^3}{kT}\frac{4\pi
a^2}{3}\int_0^{\infty}dx \frac{x^7}{1+(T/T_a)x^2}e^{-x^2}
\label{tauclass}
\end{equation}
with $\bar{\omega}^3=\omega_x\omega_y\omega_z$, while the
characteristic temperature $T_a$ is defined by $kT_a=\hbar^2/ma^2$.
The integral in (\ref{tauclass}) comes from averaging over momentum
and space the cross section multiplied by an appropriate weight
function. In the unitarity limit, (\ref{tauclass}) gives
$\tau^{-1}=4N\hbar^2\bar{\omega}^3/[15\pi (kT)^2]$. Using
(\ref{Determinant}) and (\ref{tauclass}) we can now calculate the
frequency and damping rate $\Gamma=-{\rm Im}\omega$ of the scissors
mode as a function of temperature and scattering length. In the
hydrodynamic limit $\tau\rightarrow0$, we obtain from
(\ref{Determinant})
\begin{equation}
\omega=\omega_h-i\tau\frac{2\omega_x^2\omega_y^2}{\omega_x^2+\omega_y^2}.
\label{hydrodynamic}
\end{equation}
In the collisionless limit $\tau\rightarrow\infty$, Eq.\
(\ref{Determinant}) yields
\begin{equation}
\omega=\omega_{cj}-\frac{i}{4\tau} \label{collisionless}
\end{equation}
with $j=1,2$. This result, $\Gamma=1/4\tau$, for the damping rate in
the collisionless limit, obtained by taking moments of the kinetic
equation, is in fact exact. This can be seen by treating the
collision integral as a perturbation, following the approach used in
Ref.\ \cite{kavoulak1} for the collisional relaxation of trapped
bosons above their condensation temperature.

\section{Comparison with experiment}
We now compare our calculated frequency and damping rate of the
scissors mode with the recent experiment on trapped $^6$Li
atoms~\cite{Wright}.
\begin{figure}
\includegraphics[width=\columnwidth,height=0.8\columnwidth,angle=0,clip=]{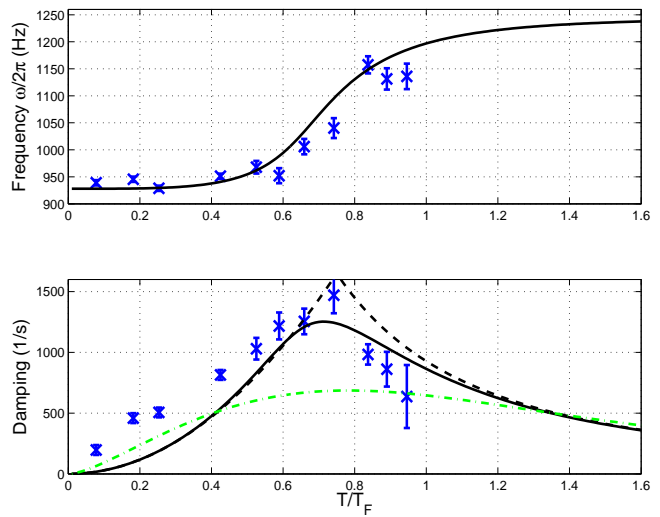} \caption{(Color online) The
frequency and damping of the scissors mode in the unitarity limit. The
$\times$'s are experimental values from \cite{Wright}. The black
lines are the theoretical values obtained from the solutions to
(\ref{Determinant}), while the dashed lines indicate the
asymptotic results (\ref{hydrodynamic}) and (\ref{collisionless}).
The green dash-dotted curve is the damping obtained using  (\ref{Emec}-\ref{dynhydro}) and (\ref{rateofloss}) with $\omega=\omega_h$.}
\label{ScissorUni}
\end{figure}
In Fig.\ \ref{ScissorUni}, we plot the observed frequency and
damping rate as a function of temperature at unitarity, $1/k_Fa=0$.
The scissors mode is excited in the $xy$ plane in a cigar-shaped
trap with $\omega_x=2\pi\times830$ Hz, $\omega_y=2\pi\times415$ Hz,
and $\omega_z=2\pi\times22$ Hz. We use the relation
$\tilde{T}\approx1.5T/T_F$ with  $kT_F=(3N)^{1/3}\hbar\bar{\omega}$
to convert the effective temperature $\tilde{T}$ obtained by fitting
the density profile to the real temperature~\cite{Wright,Kinast}.
The calculated frequency and damping from  (\ref{Determinant}) and
(\ref{tauclass}) is plotted as solid black lines. The number of
trapped atoms is taken to be $N=4\times10^5$. The dashed black lines
in Fig.\ \ref{ScissorUni} are the hydrodynamic and collisionless
limits given by (\ref{hydrodynamic}) and (\ref{collisionless}) respectively. We
see that there is good overall agreement between theory and experiment both
for the frequency and damping rate as a function of $T$. For low temperatures,  the measured damping is
somewhat larger than that obtained for our classical model. Since our model does not include Fermi blocking effects and
superfluidity, this discrepancy is not surprising.
Note that the theoretical curves contain no fitting parameters.

Due to the anharmonicity of the trap, the observed frequency in
the collisionless limit is slightly reduced from
(\ref{collisionless})~\cite{RiedlComm}, an effect which is not
included in our model, which assumes the trap to be perfectly
harmonic.

We also plot in Fig.\ \ref{ScissorUni} as a dash-dotted green line the
calculated damping rate obtained from hydrodynamics, using the
approach described below. When the damping is small, the damping rate is determined
by the ratio between the rate of loss of mechanical energy and the
energy itself. For the velocity field  given by (\ref{velfield}),
the time average of the mechanical energy, $E_{\rm mech}$, is equal
to
\begin{equation} \langle E_{\rm mech}\rangle=\frac{1}{2} \int d^3{r}mn({\bf
r})v^2({\bf r})=\frac{N}{2}kTb^2(\frac{1}{\omega_x^{2}}+\frac{1}{\omega_y^{2}}),\label{Emec}
\end{equation}
since in the classical limit the density is proportional to
$\exp(-V({\mathbf{r}})/kT)$. The time average of  the rate of loss of mechanical
energy is
$\langle\dot{E}_{\rm mech}\rangle=-2b^2\int d^3r\eta, $
from which the damping rate is obtained as
 \begin{equation}
 \Gamma=\frac{|\langle\dot{E}_{\rm mech}\rangle|}{2\langle E_{\rm mech}\rangle} .\label{dynhydro}
\end{equation}
In the unitarity limit in the classical regime, a variational calculation which is
accurate to $\lesssim1\%$ yields~\cite{Bruun75}
\begin{equation}
\eta=\frac{15}{32\sqrt{\pi}}\frac{(mkT)^{3/2}}{\hbar^2}
\end{equation}
for the viscosity. This   is independent of density and the integration must be cut off close
to the surface of the cloud where the gas is no longer
hydrodynamic~\cite{Kavoulakis}. A cut-off can also be
introduced by using the
real part of the (complex) dynamical viscosity
$\eta(\omega)=\eta/[1-i\omega\tau_\eta({\bf r})]$~\cite{Nikuni}
giving
\begin{equation}
\langle\dot{E}_{\rm mech}\rangle=-2b^2\int d^3{ r}\frac{\eta}{1+\omega^2\tau_\eta({\bf r})^2}, \label{rateofloss}
\end{equation}
where in the classical limit
\begin{equation}
\tau_{\eta}({\bf r})=\frac{\eta}{n({\bf r})kT}=\frac{4.17}{N\bar{\omega}}\left(\frac{kT}{\hbar\bar{\omega}}\right)^{2}e^{V({\bf r})/kT}.
\end{equation}
As can be seen in Fig.\ \ref{ScissorUni}, this procedure leads to results which are
qualitatively correct also in the collisionless regime. It is interesting that the expression
(\ref{rateofloss})  yields a damping rate which is exact in the collisionless limit for a trap with cylindrical symmetry
($\omega_x=\omega_y$) provided $\omega$ is set equal to its value in the collisionless limit,  $\omega=\omega_{c1}$.
For a general trap geometry, (\ref{rateofloss}) with $\omega=\omega_{c1}$ yields a
 damping rate in the collisionless limit which
differs from (\ref{collisionless}) by the factor
$8\alpha^2(1+\alpha)^{-2}(1+\alpha^2)^{-1}$ where
$\alpha=\omega_y/\omega_x$. For the trap parameters of the
experiment in Ref.\ \cite{Wright}, the factor is $\approx0.71$.

\begin{figure}
\includegraphics[width=0.9\columnwidth,
height=0.6\columnwidth,angle=0,clip=]{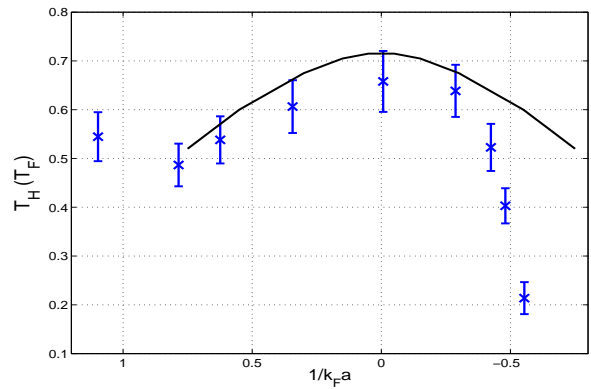} \caption{(Color online) The crossover
temperature $T_H$ between hydrodynamic and collisionless dynamics as
a function of interaction strength. The $\times$'s are experimental
values from \cite{Wright} and the solid line is obtained as described in the text.} \label{THfig}
\end{figure}

We now examine the transition between hydrodynamic and
collisionless dynamics as a function of interaction strength.
As in Ref.\ \cite{Wright}, we define the temperature $T_H$ where
the damping is maximum as marking the transition between the
hydrodynamic and collisionless regimes.
In Fig.\ \ref{THfig}, we plot as a solid line $T_H$ calculated from
(\ref{Determinant}) and (\ref{tauclass}) as a function of the
interaction strength $k_Fa$. The gas can be regarded as
hydrodynamic below and collisionless above the line.
We also plot the experimentally determined $T_H$ from
Ref.~\cite{Wright}. As expected, the transition between
collisionless and hydrodynamic behavior occurs at lower
temperatures with decreasing interaction strength. There is good
agreement between theory and experiment close to the unitarity
limit. However, the calculations do not bring out the observed surprisingly steep
decrease in $T_H$ with increasing values of $-(k_Fa)^{-1}$ on the BCS
side of the resonance. This could be due to Fermi blocking effects
 making the system less hydrodynamic.

The calculations reported here do not take into account Fermi
blocking or superfluidity. In the weak-coupling limit ($k_F|a|\ll 1$)
it is straightforward to include Fermi blocking effects  in the
collision integral as described in Ref.~\cite{Massignan}. The effect
of superfluidity was discussed in Ref.~\cite{Bruun72} for a uniform
superfluid. The viscosity, which is associated with the motion of
the normal component, decreases below the transition temperature
because of the energy gap in the spectrum of elementary excitations.
The corresponding change in the viscous relaxation rate is
numerically quite small (see Eq.\ (28) of Ref.~\cite{Bruun72}). For a
trapped superfluid the energy gap depends on position, and it is therefore
much more difficult to give a quantitative account of the
effects of superfluidity on the mode frequencies.

\section{Conclusion}
We have analyzed the scissors mode of an interacting Fermi gas in
the classical regime. By taking moments of the Boltzmann equation, we have calculated the
frequency and damping of the mode both as a function of temperature
and interaction strength. The calculation
reproduces the main features of the recent experimental findings and
can be used to identify  non-classical effects for strongly interacting
Fermi gases.

\section{Acknowledgments}
We are grateful to Allan Griffin and Stefan Riedl for very helpful
discussions.


\begin{thebibliography} {99}
\bibitem{Giorgini} S.\ Giorgini, L.\ P.\ Pitaevskii, and S.\ Stringari, arXiv:0706.3360.
\bibitem{Grimm}M.\ Bartenstein \textit{et al.},
Phys.\ Rev.\ Lett.\ \textbf{92}, 203201 (2004).
\bibitem{Kinast1}J.\ Kinast, A.\ Turlapov, and J.\ E.\ Thomas,
 Phys.\ Rev.\ Lett.\ \textbf{94}, 170404 (2005).
 \bibitem{Wright}M.\ J.\ Wright \textit{et al}., arXiv:0707.3593.
\bibitem{Massignan} P.\ Massignan, G.\ M.\ Bruun, and H.\ Smith, Phys.\ Rev.\ A
\textbf{71}, 033607 (2005).
\bibitem{Odelin}D.\ Gu\'{e}ry-Odelin and S.\ Stringari, Phys.\ Rev.\ Lett.\
\textbf{83}, 4452 (1999).
\bibitem{Bruun72} G.\ M.\ Bruun and H.\ Smith, Phys.\ Rev.\ A
\textbf{72}, 043605 (2005).
\bibitem{kavoulak1} G.\ M.\ Kavoulakis, C.\ J.\ Pethick, and H.\ Smith, {\it Phys.\ Rev.\/}
{\bf A}61, 053603 (2000).
\bibitem{Kinast}J.\ Kinast  \textit{et al}., Science \textbf{307}, 1296 (2005).
\bibitem{RiedlComm}S.\ Riedl (private communication).
\bibitem{Bruun75} G.\ M.\ Bruun and H.\ Smith, Phys.\ Rev.\ A
\textbf{75}, 043612 (2007).
\bibitem{Kavoulakis}G.\ M.\ Kavoulakis, C.\ J.\ Pethick, and H.\ Smith,
Phys.\ Rev.\ A \textbf{57}, 2938 (1998).
\bibitem{Nikuni} T.\ Nikuni  and A.\ Griffin, Phys.\ Rev.\ A
\textbf{69}, 023604 (2004).
\end{thebibliography}
\end{document}